\documentstyle[12pt]{article}
\newcommand{\re}{\par\hangindent=0.5cm\hangafter=1\noindent}

\newcommand{\gsim}{\raisebox{0.3mm}{\em $\, >$} \hspace{-3.3mm}
\raisebox{-1.8mm}{\em $\sim \,$}}
\newcommand{\bm}{\boldmath}

\newcommand{\bmr}{\mbox{\bm $r$}}
\newcommand{\bmv}{\mbox{\bm $v$}}
\newcommand{\bmF}{\mbox{\bm $F$}}

\newcommand{\bmnabla}{\mbox{\bm $\nabla$}}
\begin{document}

\baselineskip 0.84cm

\begin{center}
{\large {\bf
EARLY COSMIC FORMATION OF MASSIVE BLACK HOLES}}
\end{center}
\vspace{0.2cm}
\begin{center}
Masayuki U{\footnotesize MEMURA}\\
{\it Princeton University Observatory, Princeton, NJ 08544\\
National Astronomical Observatory, Mitaka, Tokyo 181}\\
Abraham L{\footnotesize OEB}\\
{\it Institute for Advanced Study,
Princeton, NJ 08540}
and\\
Edwin L. T{\footnotesize URNER}\\
{\it Princeton University Observatory, Princeton, NJ 08544}\\
\end{center}
\vspace{1cm}
\begin{center}
submitted to\\
{\it Astrophysical Journal Letters}
\end{center}

\baselineskip 0.64cm

\newpage

\begin{center}
{\bf ABSTRACT}
\end{center}

The evolution of nonlinear density fluctuations around the Jeans mass
shortly after cosmological recombination is analyzed using a 3D
hydrodynamics/dark--matter code. The Cosmic Background Radiation (CBR)
exerts Compton friction on free electrons due to peculiar velocities.
The dynamics therefore depends strongly on the gas ionization history.
Under a variety of ionization conditions and in systems with or without
non-baryonic components,
the baryons lose angular momentum
efficiently and collapse to form a compact optically--thick object
which would probably quickly evolve into a massive black hole.
Attention is concentrated on elucidating some of the novel physical
effects in early cosmological collapses, but
ways in which more realistic calculations might be made and in which the
scenario could be incorporated into a more complete cosmogonic model are
discussed.

\begin{center}
{\bf 1. INTRODUCTION}
\end{center}

Although it has long been known (Dicke and Peebles 1968) that the Jeans
mass in the cosmic plasma just after recombination is
$M_{Jb}=1.27\times 10^6  M_\odot \Omega_B \Omega_0^{-1/2} h^{-1}$ in
baryons (where $\Omega_B$ and $\Omega_0$ are respectively the baryonic and
the total
density parameters at present and $h$ is
the present Hubble constant in units of 100 km s$^{-1}$ Mpc$^{-1}$),
almost all recent discussions of cosmic
structure formation concentrate on much larger mass scales collapsing
at much later times.  The basic justification has been that CBR limits
on recombination era density fluctuations on mass scales of
$\gsim 6\times 10^{12}M_\odot \Omega_0^{-2} h^{-1}$,
combined with popular theoretical fluctuation power spectra,
extrapolate to quite small amplitudes at the Jeans mass scale, thus
implying that a long period of linear growth must ensue before anything
of interest occurs.  However, while not unreasonable, these considerations
are by no means compelling.  In fact, there are no significant direct
limits on the fluctuation amplitudes on small mass scales at recombination.
Moreover, theoretical knowledge of the power spectrum of initial density
fluctuations is tentative and speculative at best, and non-linear
perturbations on the Jeans mass scale at recombination are at least a
logical possibility.  In fact,
such an amplitude on small mass scales is compatible with cosmological models
involving Primordial Isocurvature Baryon (PIB) perturbations (Peebles 1987)
or topological defects (e.g. Albrecht \& Stebbins 1992).

More positively, there are several semi--empirical indications that very
early structure formation on small mass scales might have actually
occurred.  The existence of apparently high mass ($\geq 10^8 M_{\odot}$)
black holes needed to power observed quasars at redshifts up to at
least 5 (Turner 1990) could be more easily understood if some population
of far smaller mass black holes formed early enough to grow to the
required size by accretion.  The increasingly stringent limits on small
angular scale fluctuations in the CBR could be more easily accommodated
in many structure formation scenarios if the cosmic plasma were re-ionized
at redshifts 30--100, thus hiding the primordial signal
(Doroshkevich, Zel'dovich, \& Novikov 1967;
Peebles 1987; Gouda \& Sugiyama 1992).
The fact that the IGM is observed to be ionized out to redshifts of
at least 5 (Schneider, Schmidt, \& Gunn 1991)
adds plausibility to the idea.  Aside
from the possibility of re-ionization by decaying particles
(Fukugita \& Kawasaki 1992, and references therein), these
scenarios require substantial early structure formation to provide the
energy needed to ionize the IGM via stars or black hole accretion.
Another intriguing
model, recently suggested by Gnedin and Ostriker (1992), would allow the
dynamically detected dark matter in the universe to be baryonic given a
just post-recombination population of black holes producing a gamma-ray
radiation field that would modify the results of Big Bang light element
nucleosynthesis. Also, black holes could be a possible resolution
for the composition of the dark halos around galaxies
and the heating sources of stellar disks (Lacy and Ostriker 1985).
Furthermore, the existence of substantial heavy element abundances
in intergalactic clouds at redshifts above 3 (Steidel and Sargent 1988),
in intracluster gas far from the
galaxies at low redshift (Hatsukade 1989), and in Pop II stars
(da Costa 1991; Carney 1992; Suntzeff 1992)
also points to an early, pre-galactic era of structure formation
(Matsuda, Sato, \& Takeda 1969; Truran \& Cameron 1971;
Yoneyama 1972; Peebles 1974; Silk 1977, 1983;
Fall 1979; Carr, Bond, \& Arnet 1984).  Finally,
even if {\it typical} fluctuation amplitudes
on the Jeans scale were as small as usually imagined at recombination, there
would still be a small amount of mass in rare, many $\sigma$ peaks; these
unusual structures might be particularly important or interesting precisely
because they could form and influence their surroundings long before most
of the action took place.

For all of these reasons, and also simply because it has been insufficiently
explored to date, we here report a preliminary numerical investigation of
structure formation on Jeans mass scales immediately following the
recombination epoch.  Loeb (1993) recently pointed out that structure
formation at redshifts above $\sim 160$
will be strongly influenced by radiative
drag effects on the baryonic components of any fluctuations if the
gas becomes ionized or contains dust.
In particular, his analytic treatment of the evolution
of very idealized perturbations showed that the Compton drag could suppress
the centrifical barriers produced by the mutual tidal spin ups of neighboring
objects expected during gravitational structure formation
(Hoyle 1949; Peebles 1969; Ryden 1988; Heavens \& Peacock 1988).
This would make the formation of very compact systems, and in
particular black holes, much more likely (Fowler 1966; Fricke 1973).
In this Letter, we report a
preliminary numerical study of more realistic, though still quite
idealized, early structure formation using a newly developed code
(Umemura et al. 1993) devised specificly for this purpose.  At this stage
our goal is not to present a well developed theory, but rather to make an
initial survey of some of the major physical processes and of how they
affect the resulting structures.  We shall concentrate particularly on
the possibility of efficiently forming very compact objects, which might
be or rapidly evolve into black holes.

\S 2 of this {\it Letter} describes the calculations we have performed,
the physical processes included, the models considered, and the numerical
techniques employed.  The results are presented in \S 3 and conclusions
and discussion are given in \S 4.

\begin{center}
{\bf 2. FORMULATION OF THE PROBLEM}
\end{center}

The Compton friction force exerted by the Cosmic Background Radiation (CBR)
at a redshift $z$
is proportional to the electron peculiar velocity
by the factor, $\alpha(z)=\alpha_0 (1+z)^4$,
\begin{equation}
\alpha_0={4 \over 3} {\sigma_T \varepsilon_{\gamma 0} \chi_e \label{comp}
\over \mu m_p c}
\end{equation}
where $\sigma_T$ is the Thomson cross section, $\varepsilon_{\gamma 0}$ is
the present energy density of the CBR, $\chi_e$
is the ionization fraction, and $\mu$ is the molecular weight of the gas.
The corresponding equation of motion for the plasma is,
\begin{equation}
{d\bmv \over dt}=-\bmnabla \Phi - {1 \over \rho}\bmnabla p
+ \bmF\times\bmr
-\alpha (\bmv -H\bmr), \label{dvg}
\end{equation}
where $\bmv$, $\rho$ and $p$ are the gas velocity, mass density,
and pressure, $\Phi$ is the net gravitational potential
(including the dark matter), $H(t)$ is the
Hubble parameter, and $\bmF\times\bmr$ is an external tidal force.
The dark matter particles follow,
${d\bmv/dt}= -\bmnabla \Phi +\bmF\times\bmr$.
Finally, the thermal history of the gas is described by the energy equation,
\begin{equation}
{\rho \over \gamma-1}{d\over dt}\left( {P \over
\rho} \right)-{P \over \rho}
{d\rho \over dt}=-\Lambda ~~, \label{energy}
\end{equation}
where $\gamma=5/3$ is the adiabatic exponent,
and $\Lambda$ is the cooling function combining Compton,
Bremsstrahlung, recombination, and line cooling
for a primordial abundance
with hydrogen and helium (Umemura 1993 et al.).

For simplicity, we consider a quasi--spherical top--hat density perturbation
that acquires angular momentum in the $z-$direction
from an external tidal field,
$|\bmF|=0.28 (t/t_i)^{2/3}GM_T/R_i^3$, where $M_{T}=2\times10^{7}M_{\odot}$
is the total mass of the perturbation and $R_i=28.4$pc
is its initial radius.
The amplitude of the external torque is fixed so that it increases
the spin parameter of the dark matter up to a value of
$\lambda\equiv J_T|E_T|^{1/2}G^{-1}M_T^{-5/2}=0.05$ at maximum expansion,
where $E_T$ and $J_T$ are the total energy and angular momentum
(Peebles 1971, Barnes \& Efstathiou 1986, Gunn 1987,
Zurek, Quinn, \& Salmon 1988),
and it rises in time according to linear theory
(Hoyle 1949; Peebles 1969; Ryden 1988).
The initial amplitude of the perturbation shortly after the
cosmological recombination ($z=10^{3}$) is set to $\delta\rho/\rho=2$
for both the baryons and the dark matter on the corresponding mass scale.

The effects of the Compton coupling to the CBR
depend strongly on the ionization history of the
gas involved in the perturbation.
We parametrize this dependence by assuming partial ionization
of $\chi_{HII}=4\times 10^{-4}$,
$\chi_{HeII}=5\times 10^{-5}$, and
$\chi_{HeIII}=10^{-12}$ before a redshift $z_{ion}$
and complete ionization afterwards.
Different examples are tabulated
in Table 1.
Models 2, 4, \& 5 with $z_{ion}$ later than the collapsing redshift, $z_c=463$,
can be associated with internal sources of UV radiation like
massive stars; model 3 with $z_{ion}=1000$ assumes an
external source of ionizing radiation.
In all cases the Hubble constant is $H_{0}=50$km s$^{-1}$ Mpc$^{-1}$.
During the evolution of the gas cloud we calculate its radial optical
depth around the center of mass (averaged over angles),
$\tau\equiv \int \langle n_{e}\rangle_{d\Omega}\sigma_{T} dr$,
where $n_{e}\equiv \rho \chi_{e}/m_{p}$ is
the electron density. If $\tau$ exceeds
unity in a sphere of radius $r$ the Compton coupling of the gas
enclosed by this sphere to the CBR is turned--off
(in reality, the CBR affects only the outer layer of one optical depth in
thickness). The non--sphericity of the gas distribution is averaged out
because of numerical difficulties in applying the above algorithm in 3D.
If the gas cloud becomes thick with a small amount of rotation it will
eventually cool and form a quasi--spherical supermassive star which is
supported by radiation pressure and eventually collapses to
a black hole after $\sim 10^{4}{\rm yr} \times(M/10^{5}M_{\odot})^{-1}$
(Bisnovatyi--Kogan, Zeldovich \& Novikov 1967, Shapiro \& Teukolsky 1983).
The radius of the star is related
to its central temperature $T_{c}$ and mass $M_{s}$ by the relation (Wagoner
1969), $R_{s}= 0.06{\rm pc}\times
(M_{s}/10^{5}M_{\odot})^{1/2} (T_{c}/10^{4}{\rm K})^{-1}$;
the star becomes gravitationally unstable to a free--fall collapse
into a black hole when $R_{s}= 0.5 {\rm A.U.}
\times(M_s/10^{5}M_{\odot})^{3/2}$
(Shapiro \& Teukolsky 1983).
On the other hand if rotation is dominant, the cloud
would cool and form a low--entropy rotationally--supported configuration
(a supermassive disk) that could slowly shrink to
form a black hole at its center due
to angular momentum transport by viscous effects (Wagoner 1969;
Loeb \& Rasio 1993).

The nonlinear dynamics of the gas is followed
by a Smooth Particle Hydrodynamics (SPH) code that was
optimized for the present calculation
(Umemura et al. 1993).  The SPH method is combined with
an $N$-body scheme in which the gravitational interactions are
calculated by direct summations
using the special purpose processor GRAPE-1A (Fukushige et al. 1992).
The combined code was tested extensively (Umemura et al. 1993);
in particular, simulations with
several thousand particles conserve
the total energy and momentum to
an accuracy of $0.2-0.4\%$ and $2-5 \times 10^{-5}$, respectively.
In this letter we use 5000 particles that are randomly distributed
inside a sphere of radius $R_{i}$ at the initial redshift, $z=10^{3}$.
In the dark--matter cosmologies (Models 1-4) only half of the particles are
gaseous, so that the mass of a gaseous particle is 400$M_\odot$ and that
of a dark--matter particle is $7600 M_{\odot}$.
Also, the spatial resolution is limited in terms of the size of smoothed
particles. The typical scale-height of smoothed particles is about 0.08 pc
in dense regions.

\begin{center}
{\bf 3. RESULTS}
\end{center}

The top--hat perturbation prescribed in \S 2 would have reached
maximum expansion at $z=752$ and collapsed by $z=473$ if it were purely
spherical.
The actual collapse is delayed to $z_{c}=463$ due to the induced
angular momentum. After the collapse, the dark matter component undergoes
violent relaxation to form a core--halo configuration with a core radius
of 1.5 pc. Without Compton drag, the baryonic component dissipates its energy
and collapses to a disk which is stabilized by the
extended dark--matter halo.
(This mechanism is similar to that reported by Katz and Gunn (1991)).
The disk is characterized by a solid--body rotation
law inside 1.8pc with a rotation period of $4.7\times 10^4$ yr
(the age of the universe equals $1.3\times 10^{6}$ yr at $z_{c}$),
and by a flat rotation curve for $1.8<r<2.7$pc with
a rotation velocity of 235 km s$^{-1}$.

In analogy with galactic systems, it is plausible that
the collapse to a stable disk will trigger star formation activity.
In Models 2 and 4 the gas
is assumed to be entirely ionized after the disk forms (see Loeb (1993) for
the energetic requirements).
For a fully--ionized gas the Compton drag timescale is shorter than the
cosmological expansion time by $0.1\times ([1+z]/400)^{-5/2}\Omega_{0}^{1/2}$.
Therefore, the disk loses angular momentum very effectively.
Figure 1 presents the temporal evolution of the total angular
momentum for Model 2. The angular momentum of the dark matter increases
until turn around ($z=752$) due to the external tidal torque.
As the cloud collapses, angular momentum is
exchanged between dark--matter and baryonic clumps.
Although both components show rapid variations, the sum of their angular
momenta declines smoothly due to the action of the CBR on the free electrons.
Most of the angular momentum is extracted from the baryons.
Since the rotation period is shorter than the Compton drag timescale,
the gaseous disk shrinks adiabatically. Figure 2 compares
the disk without (Model 1) and with (Model 2) the CBR drag.
Between $z=395$ and $342$ the ionized disk shrinks by about a factor of 2 in
rad
 ius
while completing $\sim 8$ rotations. Figure 3 shows the radial
distributions of the mass, the centrifugal barrier radius,
the density, and the specific angular momentum of the baryons.
The centrifugal barrier radius, $j_{B}^{2}/GM(r)$, is normalized
by the smallest radius of a stable supermassive star with
a mass of $10^{5}M_{\odot}$.
The gaseous disk in Model 2 shrinks with a nearly solid body rotation due to
the potential of the dark matter core, namely
\begin{equation}
{ v_\phi^2 \over r } = { 4\pi \over 3} \rho_c r.
\end{equation}
where $\rho_{c}\approx const$ is the core density.
Therefore,
\begin{equation}
{ \dot{r} \over r } = { \dot{v_\phi} \over v_\phi} =-\alpha(t),\label{drdt}
\end{equation}
This relation holds as long as the disk is optically--thin and supported
by a uniform central density.
With a constant ionization fraction, equations (\ref{comp}) and (\ref{drdt})
yield (Loeb 1993),
\begin{equation}
r_{2}/r_1=\exp [{2\over 5}H_0^{-1}\alpha_0
\{ (1+z_2)^{5/2}-(1+z_1)^{5/2} \}].
\end{equation}
For example, $z_1=400$ and $z_2=342$ provide $r_{2}/r_{1}=0.2$ for a
fully ionized plasma.
The predicted decrease in radius is larger by a factor of $\sim 2.5$
than found in Figures
1 and 2b. The actual change is smaller because $35\%$ of
the baryonic mass becomes optically--thick by $z=342$
and the baryons share a
fraction of their angular momentum loss with the dark matter.
The specific angular momentum of the baryons follows,
\begin{equation}
{ \dot{j} \over j } = { \dot{r} \over r}+{ \dot{v_\phi} \over v_\phi}
=2{\dot{r}\over r} .
\end{equation}
For purely optically--thin baryons $j$ would have been reduced to 5\% of
its original value in the above redshift interval
but the actual reduction is
$\sim 2.5^{2}\times5\%=30\%$ as indicated by Figure 1.
The top panel of Figure 3 indicates by arrows the radii at which the optical
depth from the center equals unity.
Table 1 gives the mass of the central dense
core in the gas distribution $M_{c}$ that is
optically thick.
This core can potentially lose its
angular momentum by shedding mass
along the equatorial plane and can eventually collapse to form
a massive black hole (Bisnovatyi--Kogan, Zel'dovich \& Novikov 1967,
Loeb \& Rasio 1993).
If the ionization precedes the collapse, the dynamics of the
the gas cloud is qualitatively different. Model 3 assumes that
the ionization takes place at $z=10^{3}$. Initially the baryons
are
dragged with the CBR; although their overdensity does not grow, they lose
angular momentum effectively.
The dark matter collapses as before to form a core halo configuration.
Later on, the gas near the center falls into
the dark matter potential well (Figure 2c). This effect results from
the fact that the amount of mass enclosed within
the inner baryonic shells is increased
appreciably due to the collapse of
the dark matter (Loeb 1993). Finally, a compact quasi--spherical object
forms as shown in Fig. 3.
The inner optically--thick region of the system is likely
to form a supermassive star
with a mass of $2.7 \times 10^5 M_\odot$
which is 27\% of the total baryonic mass. This star could accrete
additional mass while cooling at the Eddington luminosity and collapsing to
form a black hole on a relatively short timescale
(Bisnovatyi--Kogan, Zel'dovich \& Novikov 1967, Loeb \& Rasio 1993).
The mass of the optically--thick core is plotted as a function of redshift
in Figure 4.

When the ionization occurs shortly after the collapse (Model 4)
the disk is subject to a self--gravitational instability due to its high
density. Consequently, it breaks into two parts with masses
$\sim 1.5 \times 10^5 M_\odot$.
The resulting dense objects are optically--thick, so that their orbital
angular momentum cannot be dissipated effectively by Compton drag.
This configuration is likely to evolve into a binary system
of supermassive stars. Dynamical friction and gravitational radiation
could later reduce the orbital angular momentum of the binary.

Finally, we simulated the evolution of a pure baryonic perturbation
(Model 5). Since $\Omega(z)=(1+[\Omega_{0}^{-1}-1]/[1+z])^{-1}$
and standard nucleosynthesis provides
$\Omega_{B_{0}}\approx 0.05 h_{50}^{-2}$ (Walker et al. 1991)
we use $\Omega_{B}\approx 1$ at the relevant redshift interval.
The ionization is assumed to occur shortly after the collapse. Due
to the high central density, the computational timestep is $\sim 10^2$ yr
and the dynamical evolution was followed only down to $z=459$.
Even at this relatively early time,
a dense compact object is found near the center
with $\tau>1$  and a mass of
$1.6 \times 10^7 M_\odot$ which is 80\% of the total
mass.
In fact, as shown in Fig. 3, this model produces a more compact and
rapidly spinning central object than any other model.  Of course,
there is no increase in the system's total angular momentum (compared to
Model 1); rather
the collapse
just brings higher specific angular momentum
material into the inner regions.  The equilibrium rotationally supported
disk requires higher angular momentum at a given radius
to support it than in Model 1 because
all of the mass is in collapsing baryons rather than largely in a more diffuse
dark matter halo.  Also, the Compton friction is
suppressed somewhat by the large optical depths implied by all baryonic mass.
Most of the Model 5 system is optically shielded immediately at the epoch of
collapse and ionization.

\begin{center}
{\bf 4. CONCLUSIONS AND DISCUSSION}
\end{center}

The primary conclusion of this investigation is that highly compact
baryonic objects, plausible progenitors of massive black holes, can
form efficiently shortly after the cosmic recombination epoch in the
collapse of density fluctuations having approximately the Jeans mass
if the gas component is re-ionized.  This conclusion holds in a reasonably
wide range of scenarios corresponding to various ionization histories
and to models with and without dominant non-baryonic dark matter components.
Compton friction of the re-ionized gas against the CBR is a key physical
process which can effectively remove angular momentum, thus facilitating
the formation of more compact objects.  The specialized 3D hydrodynamic
and $N$-body numerical code (Umemura et al. 1993) therefore allows us to
confirm the
main results and implications of 1D analytic studies of more idealized
situations (Loeb 1993).

The simulations have also revealed the rich and complex dynamical
behavior which can result from the addition of Compton drag forces
to gravitational collapses.  In cases in which the
gravitationally dominant dark matter component does not feel the drag
but the physically crucial baryonic component does, the interplay
between the two can be particularly subtle.  An example is the way
in which the rotating dark matter component can act as a reservoir of
angular momentum for the baryons which are constantly losing their spin
to the photons (see Figure 1).

The preliminary nature of this first numerical investigation must
be emphasized.  The simulations could be made much more realistic in
many ways; less symmetric initial conditions could be used, more realistic
ionization and tidal torque histories could be imposed, more natural
and chaotic "shapes" (i.e., not "top hats") could be used, interaction with
material outside the original perturbation (i.e., infall) might be
considered, star formation resulting in gas depletion and ionization
could be modeled, and so forth.  Rather than attempting a maximally
realistic calculation, we have first tried to concentrate on
elucidating the primary physical effects and their consequences.

Furthermore, for the present we have not attempted to place the study
of these early collapsing structures in the context of any more general
cosmogonic theory.  It is nevertheless clear that they are of potential
interest in a number of such contexts.  For example, structure formation
theories such as PIB (Peebles 1987) or those which invoke topological defects
will produce substantial nonlinearities on small mass scales (with possibly
non--Gaussian probabilities)
at the recombination epoch.  More generally, if the initial power spectrum of
Gaussian density fluctuations is a power law of index $ n \gsim -1$
at the recombination epoch,
then collapses
such as those studied here will be important.  Even in more conventional
cosmogonic scenarios in which most material does not find itself in a
collapsing structure until much later epochs (when the drag is no longer
important), a few very rare such objects can be expected to form
and may
be of special interest precisely because they are so unusual.

In addition to explaining the origin of the fluctuations, a more complete
theory will have to give some account of the subsequent evolution of
the compact baryonic objects (supermassive stars or disks) which result from
their collapse.  It is expected that
evolution of a convective supermassive star (Wagoner 1969) will lead to mass
shedding and loss of angular momentum (Loeb \& Rasio 1993) through marginally
stable configurations and cooling will lead to a supermassive black hole.
The radiative energy released in this process and by subsequent accretion
onto the black holes may have profound cosmological consequences including
erasure of CBR anisotropies (Doroshkevich, Zel'dovich, \& Novikov 1967;
Peebles 1987; Gouda \& Sugiyama 1992) and possibly even modification of the
light element abundances (Gnedin and Ostriker 1992).  The black holes may
grow through accretion to become the progenitors of high redshift quasars
(Turner 1991).  Star formation associated with the predicted cold, high
density gas disks could produce further ionization and significant heavy
elements at high redshift.
This could not only help explain various observations (metals in the oldest
known stars and the high redshift IGM, for example) but also might give rise
to dust at high redshift.  Such dust would further increase the coupling
to the CBR and could extend the dynamical importance of the radiative drag
to redshifts as low as 60 (Loeb 1993).

In a sense then, our most general conclusion is that the evolution of
the universe between redshifts of roughly 1000 to 100 need not have been
simple, quiet and uninteresting merely because it is (currently) so
difficult to observe and thus given little attention in most models.

\begin{center}
{\bf ACKNOWLEDGEMENTS}
\end{center}

We are grateful to J. E. Gunn, J. P. Ostriker, and D. Weinberg
for helpful discussions.
One of the authors (MU) appreciates a great deal of kind hospitality
at Princeton University Observatory.
This work was supported in part by a JSPS/NSF US--Japan Cooperative Research
Grant INT-9116745,
the Grants-in-Aid of the
Ministry of Education, Science, and Culture No. 02740131 and 03740133 (MU),
a W. M. Keck Foundation Fellowship (AL),
and NASA grants NAGW-2448 and NAGW-2173 (ELT).

\newpage
\begin{center}
TABLE 1\\[3mm]
MODEL PARAMETERS\\[3mm]
\begin{tabular}{cccccc}
\hline \hline
Model & $\Omega_D$ & $\Omega_B$ & $z_{ion}$ & $z_f$ & $M_c$
\\[2mm]
\hline
1\ldots\ldots & 0.95 & 0.05 & 0 & 395 & 0\\[2mm]
2\ldots\ldots & 0.95 & 0.05 & 400 & 342 & $3.5\times 10^5 M_\odot$ \\[2mm]
3\ldots\ldots & 0.95 & 0.05 & 1000 & 188 & $2.7\times 10^5 M_\odot$ \\[2mm]
4\ldots\ldots & 0.95 & 0.05 & 463 & 340 & $1.3\times 10^5 M_\odot$ \\[2mm]
5\ldots\ldots & 0 & 1.0 & 463 & 459 & $1.6\times 10^7 M_\odot$ \\[2mm]
\hline
\end{tabular}
\end{center}

\newpage
\begin{center}
REFERENCES
\end{center}

\re
Albrecht, A., \& Stebinns, A. 1992, Phys. Lev. Lett., 69, 2615
\re
Barnes, J., \& Efstathiou, G. 1987, ApJ, 319, 575
\re
Bisnovatyi-Kogan, G. S., Zel'dovich, Ya. B., \& Novikov, I. D. 1967,
Sov. Astr., 11, 419
\re
Carney, B. W. 1992, in IAU Symposium
149, The Stellar Populations of Galaxies, ed.
B. Barbuy \& A. Renzini (Dordrecht: Kluwer), 15
\re
Carr, B. J., Bond, J. R., \& Arnet, W. D. 1984, ApJ, 277, 445
\re
da Costa, G. S. 1991, in IAU Symposium
148, The Magellanic Clouds, ed.
R. Haynes \& D. Milne (Dordrecht: Kluwer), 183
\re
Dicke, R. H., \& Peebles, P. J. E. 1968, ApJ, 154, 891
\re
Doroshkevich, A. G., Zel'dovich, Ya. B., \& Novikov, I. D. 1967,
Sov. Astr. 11, 231
\re
Fall, S. M. 1979, Rev. Mod. Phys., 51, 21
\re
Fowler, W. A. 1966, Proc. Internat. School Phys. Enrico Fermi, Vol. 35
(New York; Academic Press)
\re
Fricke, K. J. 1973, ApJ, 183, 941
\re
Fukugita, M., \& Kawasaki, M. 1992, ApJ, 402, 58
\re
Fukushige, T., Ito, T., Makino, J., Ebisuzaki, T., Sugimoto, D.,
and Umemura, M. 1991, PASJ, 43, 841
\re
Gnedin, N. Yu, \& Ostriker, J. P. 1992, ApJ, 400, 1
\re
Gouda, N., \& Sugiyama, N. 1992, ApJ, 395, L59
\re
Gunn, J. E. 1987, in The Galaxy, ed. G. Gilmore, \& B. Carswell
(Dordrecht: Reidel), p. 413
\re
Hatsukade, I. 1989, PhD Thesis, Miyazaki University, Japan
\re
Heavens, A., \& Peacock, J. 1988, MNRAS, 232, 339
\re
Hoyle, F. 1949, in Problems of Cosmological Aerodynamics,
International Union of Theoretical and Applied Mechanics and
IAU, ed. J. M. Burgers \& H. C. van de Hulst (Ohio), 195
\re
Katz, N., \& Gunn, J. 1991, ApJ, 377, 365
\re
Lacy, C. G., \& Ostriker, J. P. 1985, ApJ, 299, 633
\re
Loeb, A. 1993, ApJ, 403, 542
\re
Loeb, A., \& Rasio, F. 1993, in preparation
\re
Matsuda, T., Sato, H., \& Takeda, H. 1969, Progr. Theor. Phys., 42, 219
\re
Peebles, P. J. E. 1969, ApJ 155, 393
\re
------------. 1971, A\&A, 11, 377
\re
------------. 1974, ApJ, 189, L51
\re
------------. 1987, ApJ, 315, L73
\re
Ryden, B. S. 1988, ApJ, 333, 78
\re
Schneider, D. P., Schmidt, M., \& Gunn, J. E. 1991, AJ, 102, 837
\re
Shapiro, S. L., \& Teukolsky, S. A. 1983, Black Holes, White Dwarfs,
and Neutron Stars (New York; Wiley)
\re
Silk, J. 1977, ApJ, 211, 638
\re
------------. 1983, MNRAS, 205, 705
\re
Steidel, C. C., \& Sargent, W. L. W. 1988, ApJ, 333, L5
\re
Suntzeff, N. B. 1992, in IAU Symposium
149, The Stellar Populations of Galaxies, ed.
B. Barbuy \& A. Renzini (Dordrecht: Kluwer), 23
\re
Turner, E. L. 1991, AJ, 101, 5
\re
Umemura, M. 1993, ApJ, 406, in press
\re
Umemura, M., Fukushige, T., Makino, J., Ebisuzaki, T., Sugimoto, D.,
Turner, E. L., \& Loeb, A. 1993, PASJ, in press
\re
Wagoner, R. V. 1969, ARAA, 7, 553
\re
Walker, T., Steigman, G., Schramm, D. N., Olive, K. A., \& Kang, H.S. 1991,
ApJ, 376, 51
\re
Yoneyama, T. 1972, PASJ, 24, 87
\re
Zurek, W. H., Quinn, P. J., \& Salmon, J.K. 1988, ApJ, 330, 519

\newpage
\begin{center}
{\bf FIGURE CAPTIONS}
\end{center}
\re
FIG. 1 --- Angular momenta versus redshift
for Model 2.
The upper panel shows the total (dotted)
and the dark--matter (solid) angular momenta
after the re--ionization redshift $z_{ion}=400$. The lower panel shows
the angular momentum of the baryonic component.
\re
FIG. 2 --- The spatial distributions of the dark--matter and
the baryons at the final stages of the calculation, namely:
{\it a)} $z=395$
for Model 1,
{\it b)} $z=342$ for Model 2, and {\it c)} $z=188$ for Model 3.
The box size is 10 pc.
\re
FIG. 3 --- Final radial distribution
(around the center of mass) of the gas properties.
Top panel shows the baryonic mass within a radius $r$.
The mass unit is $10^5 M_\odot$ for Models 1, 2, and 3, and
$2\times 10^6 M_\odot$ for Model 5.
The short arrows in this panel show the radius at which the optical depth
equals unity.
The second panel shows
the ratio of the centrifugal barrier radius
to the radius of a supermassive star of the mass enclosed within that radius
at the onset of gravitational instability $R_{s}\equiv
0.5 {\rm A.U.}
\times(M_s/10^{5}M_{\odot})^{3/2}$
(Shapiro \& Teukolsky 1983).
The third panel
shows the mass density, and the  bottom panel presents
the specific angular momentum of the baryons in the $z-$direction.
The unit of specific
angular momentum is pc$^2$ yr$^{-1}$.
The various lines refer to the final time of Models 1 (solid), 2 (dot-dashed),
3 (dotted), and 5 (dashed).
\re
FIG. 4 --- Mass of the optically--thick core as a function of redshift
for Model 3.
\end{document}